\documentclass{aa}
\usepackage{graphicx}
\usepackage{subcaption}
\usepackage{txfonts}
\usepackage{natbib}
\usepackage{placeins}
\usepackage{multirow}
\usepackage{hyperref}
\usepackage{siunitx}
\usepackage{orcidlink}
\bibpunct{(}{)}{;}{a}{}{,} 
\usepackage{booktabs}
\usepackage{enumitem} 

\newcommand\dfdw{{\ensuremath{\emph{dfdw}}}}
\newcommand\dw{{\ensuremath{dw}}}
\newcommand\df{{\ensuremath{\emph{df}}}}
\newcommand\dm{{\ensuremath{\emph{dm}}}}
\newcommand\dt{{\ensuremath{dt}}}
\newcommand{\fwp}{flux-wavelength plot}

\newcommand{\Autoref}[1]{%
  \begingroup
  \def\sectionautorefname{Section}%
  \def\subsectionautorefname{Subsection}%
  \def\subsubsectionautorefname{Subsection}%
  \def\figureautorefname{Figure}%
  \def\tableautorefname{Table}%
  \def\equationautorefname{Equation}%
  \def\itemautorefname{Item}%
  \def\appendixautorefname{Appendix}%
  \autoref{#1}%
  \endgroup
}

\begin{document}

    \title{On the performance of pre-trained vision transformers for supernova spectral classification using different spectral representations}
    \titlerunning{Vision transformers for SN spectra}
    
\author{
J. Serrano Bell\inst{1,2}\orcidlink{0000-0002-8397-557X}\and
P. G\'alvez Molina\inst{3}\orcidlink{0009-0006-0459-6923}\and
V. Contreras Rojas\inst{4}\orcidlink{0009-0008-6278-0619}\and
W. Fox Fortino\inst{3}\orcidlink{0000-0001-7559-7890}\and
M. Rojas\inst{5} \orcidlink{0009-0005-9943-9084}\and
P. Protopapas\inst{6}\orcidlink{0000-0002-8178-8463}\and
F. Bianco\inst{3,7,8,9}\orcidlink{0000-0003-1953-8727}}

\institute{
\inst{1} Instituto de Ciencias Físicas (ICIFI-CONICET), ECyT-UNSAM, Campus Miguelete, 25 de Mayo y Francia, (1650) Buenos Aires, Argentina \\
\inst{2} Instituto Tecnológico de Buenos Aires (ITBA), Buenos Aires C1437, Argentina \\
\inst{3} University of Delaware, Department of Physics and Astronomy, Bartol Research Institute, Newark, DE, USA \\
\inst{4} Departamento de Astronomía, Universidad de La Serena, Raul Bitran 1720256, La Serena, Chile \\
\inst{5} Electronics Department, Federico Santa Maria Technical University, Valparaiso, CL \\
\inst{6} John A. Paulson School of Engineering and Applied Sciences, Harvard University, Cambridge, MA, USA \\
\inst{7} Joseph R. Biden, Jr. School of Public Policy and Administration, University of Delaware, Newark, DE 19716, USA \\ 
\inst{8} Vera C. Rubin Observatory, Tucson, AZ, USA \\
\inst{9} Data Science Institute, University of Delaware, Newark, DE 19716, USA \\
}

\abstract{The spectroscopic classification of supernovae is a key component of time-domain astronomy and plays an important role in the identification of Type Ia events for cosmological applications. In recent years, the field has seen a proliferation of automated, neural-network-based approaches, reflecting sustained methodological interest in classification tools that are accurate, robust, and scalable to the diversity and volume of spectra produced by modern surveys. In this work, we explore the use of Transformer-based vision models for supernova spectral classification, focusing on how different visual representations of spectra and fine-tuning strategies affect classification performance when image-based architectures are applied to intrinsically one-dimensional data. We consider a dataset of 4,011 supernova spectra, augmented and rebalanced into three classes: normal Type Ia, other Type Ia subtypes, and core-collapse supernovae. Spectra are encoded as line-plot images using direct flux-wavelength visualizations as well as alternative flux-difference versus wavelength-difference heatmap representations (\dfdw) designed to emphasize differential spectral structure. We evaluate several pre-trained Transformer architectures, including isotropic Vision Transformers (ViT), hierarchical Swin Transformers, and a self-supervised DINOv3-pretrained ViT, and examine the effects of fine-tuning depth, visual representation, and hyperparameter choices. We find that a plain flux-wavelength line plot outperforms the purpose-built \dfdw\ maps on average, though log-scale \dfdw\ maps remain competitive for specific architecture and fine-tuning combinations. On the held-out test set, our best model (\texttt{vitb-p16}) reaches a macro-$F1$ score of 86.3\%, with per-class $F1$ scores of 94\% for normal Ia, 77\% for other Ia subtypes, and 88\% for core-collapse supernovae; we note that on validation the top-ranked configuration was instead \texttt{swinv2t-p4} with the log-scaled 50$\times$50 \dfdw\ map (90.0\%), highlighting a modest validation-to-test gap. The dominant misclassification arises from confusion between Ia-91T and normal Ia spectra, a phase-dependent effect consistent with the known spectral evolution of this subtype. Our work provides an exploratory, methodological comparison of pre-trained vision-Transformer architectures and spectral encodings, identifying \texttt{vitb-p16} with the flux-wavelength representation as the most promising configuration among those tested. Our results suggest that pre-trained Transformer-based vision models can provide competitive performance for automated single-spectrum supernova classification in the context of published automated-classifier metrics.}

\maketitle
\nolinenumbers
                      
\section{Introduction}
\label{sec:introduction}

Supernovae (SNe) play a fundamental role in astrophysics, acting as key drivers of chemical enrichment, feedback in galaxy evolution, and as cosmological distance indicators. Their classification has traditionally relied on spectroscopic features, most notably the presence or absence of hydrogen and helium lines, as well as characteristic absorption features such as Si \textsc{ii} $\lambda$6355~$\AA$ in SNe Ia. The original spectroscopic classification scheme introduced by \citet{1941Minkowski} and later refined by \citet{1997Filippenko} has proven remarkably robust, forming the backbone of modern transient astronomy.

SNe Ia constitute a particularly important subclass due to their use as standardizable candles in cosmological studies \citep{1984SvA....28..658P, phillips1993absolute, riess_precise_1996, tripp_using_1997, 1997AAS...191.8504P, 1998Riess, 1999Perlmutter}. However, it is well established that SNe Ia exhibit significant spectroscopic and photometric diversity, leading to the identification of multiple subtypes and peculiar events \citep[e.g.,][]{1992Phillips,1997Filippenko,2001Li,2006Benneti}. This intrinsic heterogeneity and the time-evolving nature of spectra complicate both the physical interpretation and the automated identification in large samples.

The rapid growth of wide-field time-domain surveys will substantially increase the transient discovery rate, while spectroscopic follow-up resources remain limited \citep{2019Bellm, 2019Ivezic}. Next-generation spectroscopic facilities may help close this gap. For example, the proposed Wide-field Spectroscopic Telescope \citep[WST;][]{2026Bacon} is designed to combine a 12-m aperture with a multi-object spectrograph capable of simultaneously observing 30,000 targets at low resolution and 2,000 at high resolution. Although these capabilities are not specific to transient classification, they illustrate the scale at which future spectroscopic surveys may operate. As follow-up expands to larger transient samples, scalable and automated methods will become increasingly valuable for processing spectra efficiently and consistently.

The first spectroscopic classification approaches beyond pure visual inspection were based on template matching and cross-correlation methods (\citealt{2005Howell,2007Blondin}). These methods continue to be widely used. Yet, while spectroscopic classification of an individual transient can be performed effectively by template matching and cross-correlation tools such as SNID \citep{2007Blondin,2025Magill,Stoppa2026} or Superfit/NGSF-based methods \citep{2005Howell, 2022TNSAN.191....1G}, these methods require human supervision and may incur significant computational costs. For example, the computational cost of SNID is low per classification (6\,s × (CPU/2.86 GHz) × (Ntemplates/1000), \citealt{Blondin2012}) 
but scales linearly with the number of templates in the reference library.  Template methods based on cross-correlation techniques or minimization algorithms require user input and supervision to derive accurate classifications and mitigate spurious correlations, which would occur more frequently with increasingly larger template banks, and manage hyperparameters individually for each spectrum, so they are not and cannot be completely automated. 

Conversely, automated, neural-network based classifiers take advantage of amortized inference, shifting most of the computational cost to training.
For a more detailed discussion of the motivations for introducing machine learning level automation in the spectral classification workflow, we refer the reader to, e.g., \citealt{2019DASH, Sharma2025, 2025FoxFortino, fortino2026low, williamson2019optimal, de2023graph}.

Machine-learning techniques for the classification of SNe were introduced for photometric classification based on light curves and hand-crafted features  \citep[e.g.,][]{2016Lochner}. Deep-learning methods have been applied to both light curves  \citep[e.g.,][]{2016Moller} and spectra.
In particular, several neural-network-based approaches have been developed for automated spectroscopic supernova classification. Beginning with DASH \citep{2019DASH}, and followed by SNIa-score \citep{2021Fremling}, CCSNscore \citep{Sharma2025}, ABC-SN \citep{2025FoxFortino}, and SpectraNet \citep{xu2026applecider}, among others. These methods employ a range of deep-learning architectures, including convolutional and recurrent neural networks designed to classify or score supernova spectra, as well as more recent approaches based on attention and generative architectures. This growing body of work demonstrates that automated spectroscopic classification is an active area of research.

Our work, however, differs from these examples in a key aspect: we are not developing a new model, but rather we are exploring the potential of pretrained foundational models for the spectral classification task. Foundation models are developed and trained leveraging massive data and computational resources, typically well beyond the reach of academic research. But there is a significant domain shift in their application to astronomical data due to the inherently different nature of astronomical images, time series, and spectra compared to the real-world data these models are trained on \citep[e.g.,][]{wang2026eaa}.

Examples of works aimed at assessing the capabilities of foundation models for astrophysical tasks include, e.g.,  \citet{rouhbakhshmeghrazi2025exploring} and \citet{ 2025A&A...693A.232V} in computer vision and \citet{2025AJ....170..120C} and \citet{li2025starembed} in time domain astronomy, among others, showing that this is also a growing area of research interest.

Many deep-learning applications to SN spectra have relied on convolutional neural networks (CNNs), which are highly effective at extracting local spectral features such as absorption and emission lines \citep[e.g.,][]{2019DASH,xu2026applecider}, with recurrent architectures (e.g., bidirectional LSTMs) also being applied to sequence-level tasks such as Ia versus non-Ia scoring \citep{2021Fremling}. However, CNNs exhibit a locality bias that makes modeling long-range dependencies across the full wavelength range challenging, often requiring substantially deeper architectures or additional mechanisms \citep{2017Vaswani,2020ViT,2023Khan}.

Transformers, originally introduced for natural language processing \citep{2017Vaswani}, address this limitation through self-attention mechanisms that enable interactions between all elements of an input sequence. Their adaptation to computer vision gave rise to Vision Transformers (ViTs), which represent images as sequences of patch embeddings and have demonstrated competitive performance in image classification and representation learning tasks \citep{2020ViT}.

Several architectural variants have since been proposed to improve efficiency and representation learning. Hierarchical models such as the Swin Transformer \citep{2022Liu} introduce shifted window-based self-attention to better capture multi-scale structures while reducing computational cost, whereas self-supervised frameworks such as DINO \citep{2025Simeoni} leverage knowledge distillation without labels to learn robust visual representations. 

In recent years, Transformer-based architectures have begun to be explored in astronomical contexts, including transient classification, with encouraging results for light curves \citep[e.g.,][]{2023Allam,2023Donoso-Oliva,2024Cabrera-vives,2025Moreno-Cartagena} and spectra \citep{2025FoxFortino,2025Strano}, including stellar parameter inference \citep{lu2026spectra}. \citet{2025Strano} established that pre-trained ViTs can classify large samples of stellar and galaxy spectra from SDSS/LAMOST. 

In our work, we explore the application of pre-trained Vision Transformers to astrophysical transients with the aim of providing an initial assessment of their competitiveness with state-of-the-art automated SN classifiers and comparing architectures, fine-tuning strategies, and spectral representations under a common protocol to provide guidance on future research directions for the refinement of their application. The present work offers two novel approaches:  (i) we systematically compare visual representations---direct flux--wavelength line plots versus \dfdw\ difference maps---rather than adopting a single fixed encoding; and (ii) we evaluate several ViT architectures (isotropic ViT, hierarchical Swin, and a self-supervised DINOv3 backbone) across fine-tuning depths.

The choice of visual representation is particularly relevant when applying vision-based architectures to spectroscopy. SN spectra are fundamentally one-dimensional signals, encoding flux as a function of wavelength, but they can be mapped into image-like formats. Previous studies have explored representing spectra or other sequential astronomical data in formats amenable to image-based learning, such as stacking subsets of the spectral sequence at finite wavelength intervals leading to a 2D representation \citep{2019DASH}, converting one-dimensional spectral data into lightweight grayscale images for machine learning benchmarks \citep{2026Astrospectra}, or applying deep-learning architectures such as autoencoders to extract salient features from time-domain data \citep[e.g.,][]{2021Villar}. Nevertheless, the impact of different visual encodings on Transformer-based spectral classification remains insufficiently explored.
Here, we investigate two complementary visual representations of SN spectra. The first representation consists of one-dimensional spectra encoded as images, preserving the wavelength ordering and relative flux information. The second representation employs two-dimensional flux-difference vs. wavelength-difference maps, (\dfdw , a modification of the $dmdt$ representation introduced for light curves in \citealt{2011BASI...39..387M}), designed to emphasize differential spectral features by encoding flux variations as a function of wavelength and local spectral gradients. This approach enhances contrast in line-dominated regions and highlights subtle spectral differences that are often critical for discriminating between SNe subtypes.

We frame the classification task into three classes: normal Type Ia (Ia--norm), other Type Ia (Other Ia) subtypes, and core-collapse (CC) supernovae. This simplified taxonomy offers practical advantages: it provides a tractable common comparison given the strong subtype imbalance and the small number of independent objects in several subtypes (\Autoref{subsec:balancing}). 
This taxonomy, and the models trained on it, are not intended as a proposed replacement for existing operational classification pipelines. Rather, it serves as a controlled testbed for a proof-of-concept study of the methodological potential of off-the-shelf, pre-trained vision Transformers and of alternative visual encodings of spectra. 

While significantly simplified compared to the highly complex SN taxonomy \citep{Gal_Yam_2017}, this classification task retains a cosmologically relevant distinction between normal Type Ia and everything else and discrimination between the two core astrophysical explosions channels: compact objects and massive stars. Simplified taxonomies are commonly used in the development of classifiers. For example, \citealt{2021Fremling} develops a binary classifier designed to classify SN Ia with high purity. The more recently developed \citealt{xu2026applecider} also introduces a simplified taxonomy in three SN classes (SN Ia, SN Ibc, SNII, in addition to three non-SN classes CV, AGN, and TDE). Additionally, this taxonomy reflects early branching of hierarchical classification schema for SNe proposed, for example, in \citealt{2023arXiv231202266V}. 

The use of a non-standard taxonomy complicates comparison between classifiers' performance. Nevertheless, the classification schema allows a common comparison among the models tested here and enables a direct comparison with precursor models trained on the same datasets under a restricted common taxonomy. Published automated-classifier results provide context on overlapping classes, although such comparisons remain subject to differences in data and evaluation protocols.

This paper is organized into the following sections: \Autoref{sec:data_setgen} describes the construction of the dataset, including the labeling strategy, class balancing, and data augmentation procedures. \Autoref{subsec:data_visualization} presents the two visual representations of the spectra considered in this work, namely the direct spectral images and the \dfdw\ map encoding. The experimental setup is described in \Autoref{sec:experimental_setup}, where we introduce the Transformer-based models, the training and evaluation strategy, and the different fine-tuning configurations explored. The results and their discussion are presented in \Autoref{subseec:finalmodel} and \Autoref{sec:discussion}, respectively. Finally, \Autoref{sec:conclusion} summarizes the main conclusions of this work. All data products and code necessary to reproduce the results presented in this work are publicly available in a GitHub repository\footnote{\url{https://github.com/juanserrano90/codelatam}}.

\section{Data selection and preparation}
\label{sec:data_setgen}

\subsection{Labeling and balancing the dataset}
\label{subsec:balancing}
The SNe dataset presented by \citet{2025FoxFortino} was used as the basis for this study. This dataset is itself a modification of the templates originally developed for the Supernova Identification algorithm (SNID\footnote{\url{https://people.lam.fr/blondin.stephane/software/snid/}}; \citealt{2007Blondin})  and of its updated version used to train the DASH \citep{2019DASH}, the first deep-learning based SN spectra classifier. In addition to the core SNID templates, the dataset includes several updates and extensions: a catalog of stripped-envelope supernovae (SESNe) compiled from \citet{LiuModjaz2015}, \citet{Modjaz2014, Modjaz2016}, and \citet{Liu2016}, which were incorporated into SNID after the original DASH training, as well as spectra from the Berkeley SN Ia Program \citep[BSNIP;][]{Silverman2012}. The SNID templates themselves were originally assembled from multiple sources, including the SUSPECT archive \citep{YaronGalYam2012} and the CfA Supernova Archive and Program \citep{Matheson2008,Blondin2012}.

\begin{figure}
    \centering
    \includegraphics[width=\linewidth]{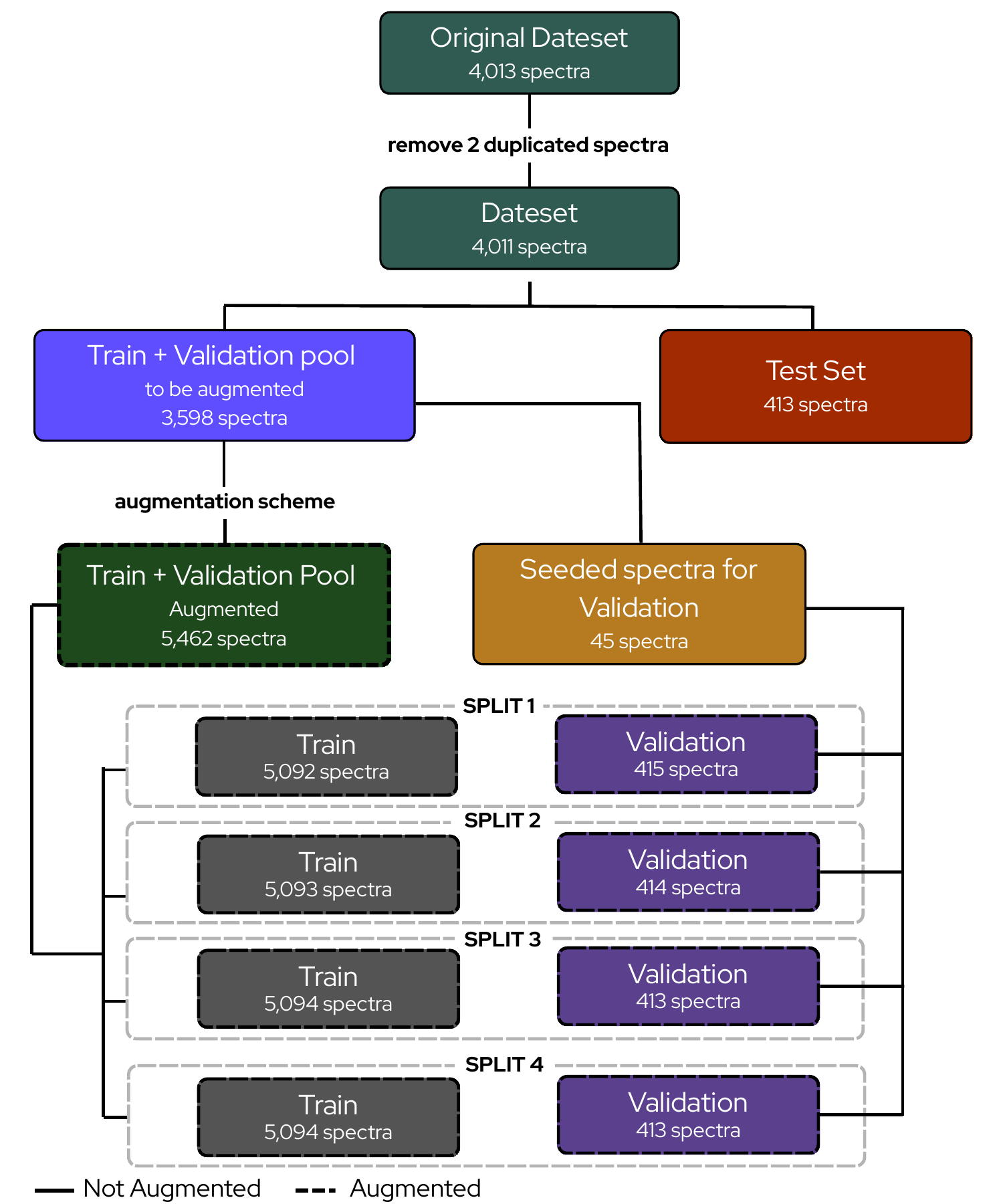}
    \caption{Schematic representation of the dataset splitting strategy. The four splits enclosed by the light gray dashed box are all drawn from the augmented training/validation pool after the Test set (red box) has been assembled. In each split, the validation set includes randomly drawn augmented spectra and the same seeded, non-augmented 45 spectra (shown in the yellow box). The number of spectra in the training, validation, and test sets for each split is indicated in the figure. The detailed subtype distributions for the spectra in each set are provided in the Appendix in \Autoref{tab:split1}--\Autoref{tab:split4}.}
    \label{fig:split-schematic}
\end{figure}

The dataset consists of 4,013 SN spectra, each associated with a spectral phase and categorized by both a ``main type'' and a ``subtype''. Each spectrum consists of 1,024 wavelength bins spanning the interval [2,501.69 -- 9,993.24]~\AA. Notably, this is a heterogeneous dataset comprised of spectra collected at different facilities, with different observational setups and signal-to-noise ratios (SNR). Furthermore, the spectra were collected at different phases in the evolution of each SN, between -20.0 days and 50.0 days. Two duplicated spectra present in the original set were dropped (LSQ14efd at a spectral phase of 31.93 days and SN 2004fe at a spectral phase of -6.78 days) for a total of 4,011 distinct spectra, as noted on the schematic in \autoref{fig:split-schematic}. This heterogeneous dataset provides instrumental diversity, but does not establish robustness or transferability to poorly represented noise regimes; we note in fact that this dataset is a curated compilation of relatively high SNR spectra, originally assembled for template matching purposes, and does not fully sample the lower SNR regime characteristic of faint transients from current and future wide-field surveys. The SNR distribution of the dataset is discussed in detail in \citealt{fortino2026low}. The data were preprocessed through the steps described in \citealt{2007Blondin}, namely: continuum subtraction, ablation, and binning on an evenly sampled wavelength grid in log-wavelength space. This is a common pre-processing schema adopted in many other works including \citealt{2019DASH,2024MLS&T...5d5069Z,2025FoxFortino}. Additionally, the spectra were already redshift-corrected to rest-frame.

The spectra were originally categorized by SNID into one of four main types: Ia, Ib, Ic, and II. In addition, the dataset includes 17 subtypes: Ia-norm, Ia-91T, Ia-csm, Ia-91bg, Ib-norm, Ia-pec, Ic-norm, IIP, Iax, IIL, IIb, II-pec, Ic-broad, Ic-pec, IIn, Ibn and Ib-pec. This detailed categorization is too complex for our study, and the classification of SN spectra by subtype remains an active topic of discussion in the field \citep[e.g.,][]{2025Kankare, 2025FoxFortino}. Subtype classes are not always sharply defined and may depend on different observational criteria (e.g., spectral phase), which can introduce ambiguity and label noise in supervised learning. Furthermore, as shown in \Autoref{tab:full-table-data-set}, the dataset is highly imbalanced at the subtype level: the Ia-norm subtype alone accounts for 52.71\% of all available spectra, while several subtypes (Ia-csm, IIL, IIn, Ib-pec, Ibn and Ic-pec) each represent less than 1\%.

To mitigate this imbalance, the classification task is restricted to distinguishing between type SN Ia-normal, all other SN Ia subtypes, and all core-collapse types, labeled as classes 0 (or Ia-norm), 1 (or Ia-other), and 2 (or CC), respectively. This reduces the complexity of the classification task, allowing us to better assess the impact of data encoding, training, and architectural choices, while remaining physically meaningful, as the identification of Type Ia-normal SNe is particularly relevant for cosmological studies \cite[][and references therein]{ishida2019machine, 2025Hayes, 2025A&A...694A.130M}, and the separation of CC from SN Ia classes reflects a distinction between massive star and compact object progenitors. However, Ia-other and CC are heterogeneous groups, so aggregate performance does not establish reliable discrimination of its constituent subtypes. Under this scheme, class 0 represents 52.71\% of the original dataset, class 1 represents 19.17\%, and class 2 represents 28.12\%.

\begin{table}[!htbp]
\caption{Dataset composition by supernova class and subtype.}
\label{tab:full-table-data-set}
\centering
\footnotesize
\setlength{\tabcolsep}{2.5pt}
\begin{tabular}{@{}llrrrr@{}}
\hline\hline
\shortstack{Class} & \shortstack{SN Subtype} & \shortstack{N° of SNe} & \shortstack{N° spectra \\ (original set)} & \shortstack{N° spectra \\ (augmented set)} & Copies \\
\hline
0 & {\bf Ia-norm} & {\bf 245} & {\bf 2,114} & {\bf 2,114} & {\bf 0} \\
\hline
\multirow{5}{*}{1} & Ia-91T  & 27 & 348  & 386  & 38  \\
                   & Ia-91bg & 28 & 232  & 385  & 153 \\
                   & Ia-csm  & 2  & 16   & 356  & 340 \\
                   & Ia-pec  & 6  & 111  & 353  & 242 \\
                   & Iax     & 3  & 62   & 381  & 319 \\
\hline
                   & {\bf Ia-other} & {\bf 66} & {\bf 769} & {\bf 1,861} & {\bf 1,092} \\
\hline
\multirow{11}{*}{2} & II-pec   & 1  & 47  & 144 & 97  \\
                    & IIL      & 2  & 10  & 148 & 138 \\
                    & IIP      & 6  & 104 & 166 & 62  \\
                    & IIb      & 19 & 232 & 232 & 0   \\
                    & IIn      & 2  & 22  & 147 & 125 \\
                    & Ib-norm  & 22 & 211 & 211 & 0   \\
                    & Ib-pec   & 2  & 12  & 149 & 137 \\
                    & Ibn      & 3  & 27  & 166 & 139 \\
                    & Ic-broad & 23 & 228 & 228 & 0   \\
                    & Ic-norm  & 20 & 204 & 204 & 0   \\
                    & Ic-pec   & 2  & 31  & 150 & 119 \\
\hline
                    & {\bf CC} & {\bf 102} & {\bf 1,128} & {\bf 1,945} & {\bf 817} \\
\hline
\multicolumn{2}{l}{\bf Total} & {\bf 413} & {\bf 4,011} & {\bf 5,920} & {\bf 1,909} \\
\hline
\end{tabular}
\tablefoot{SNe counts unique objects. Original and Augmented count spectra before and after augmentation, respectively. Copies counts the spectra added by augmentation.}
\end{table}

To measure the robustness of our model performance we generated multiple training-validation-test splits, aiming for 0.8-0.1-0.1 size ratios, respectively. However, the severe imbalance of the dataset requires the implementation of specific augmentation and splitting strategies, and while we perform the classification in three classes, we still need to ensure that the 17 subtypes are distributed as evenly as possible across all the sets. Re-balancing by up-sampling was applied on the training sets to avoid introducing bias into the network, while the test sets were not rebalanced, and thus remained representative of the breakdown in the original data, and we used the macro-$F1$ score for model evaluation, which is robust to class imbalance \citep{2020arXiv200805756G}.\footnote{We note that the original data breakdown is itself biased by data collection practices, and not representative of the rate at which different SN subtypes occur or can be detected.} 

\Autoref{fig:split-schematic} shows our data preparation workflow. The first set to be assembled was the test set.
The sets were chosen by iterating random draws until a configuration was found which satisfied a number of requirements.  
The split has to be performed at the supernova level rather than the spectrum level to avoid information leakage \citep{2025FoxFortino}.  That is, all spectra belonging to a given SN were kept together so that no SN appears in both the test set and the training/validation pool. The following additional constraints were applied: \begin{itemize}
    \item We aimed for the test set to contain $\approx10\%$ of the data. 
    \item We required class 1 to be represented by at least 3 subtypes and class 2 by at least 6 subtypes.
    
    \item To prevent a single object from dominating the test set, no single SN could contribute more than 20\% of the target spectrum count for classes 0 and 2, or more than 30\% for class 1. 
\end{itemize} 
Subtypes with fewer than three unique objects (Ia-csm, Ib-pec, Ic-pec, IIL, IIn) were excluded from the test set, given the impossibility of having at least one object in training, validation, and test. Instead, we saved them to introduce (seeded) one of the objects in every validation split as shown in \Autoref{fig:split-schematic}.  This includes SN 2002ic, SN 2008A, SN ls03D3bb, SN 2002ao, SN 2005la, SN 2005ek, SN 1992H, SN 1979C and SN 1996L (totaling 45 spectra). Note that all subtype II-pec spectra belong to the same object (SN 1987A, \citealt{arnett1989supernova}) and in that case they were all placed in the training set.

The pool of remaining spectra was augmented following the strategies described in \Autoref{subsec:dataaug}, the augmented dataset was then used to generate four train-validation splits. As with the test set, the splits were performed at the supernova level so that all spectra from a given object remained together in either training or validation. Random draws are performed until splits with the following constraints are obtained:
\begin{itemize}
    \item At the spectra level, each validation set comprises $\approx 10\%$ of the total spectra, corresponding to 211, 77, and 113 spectra for classes 0, 1, and 2, respectively, with a tolerance of $\pm$5 spectra for classes 0 and 2 and $\pm$4 for class 1. 
    (At the SN level, the targets were $\approx24$, $\approx7$, and $\approx10$ unique supernovae per class.)
    
        \item No object could contribute more than 20\% of the target spectrum count for classes 0 and 2, or 30\% for class 1.
        \item Class 1 is represented by at least 5 subtypes and class 2 by at least 10, ensuring broad subtype coverage across all four validation sets.
        
    \end{itemize}

To identify the optimal composition, candidate validation sets were generated across a range of random seeds and scored jointly on two criteria: pairwise spectral overlap between splits, and subtype imbalance across the four sets. The four splits minimizing the combined score were retained. The size of the validation set (10\%) was chosen empirically as it yielded the greatest diversity across the four splits while maintaining adequate subtype representation. For the subtypes with only two objects, in each generated candidate split we seeded one object into the training set and the other in validation.

The actual number of spectra and unique SN per class and subtype in each of the splits are shown in \Autoref{apx:spectra-distribution-splits}. For each experiment described in \Autoref{sec:experimental_setup} and for the final hyperparameter selection (\Autoref{subsec:hyperparam_optimization}), the metrics reported are computed on three training-validation-test splits (\Autoref{tab:split1}, \Autoref{tab:split2}, and \Autoref{tab:split3}), and the split of \Autoref{tab:split4} is reserved for the final training and validation of the optimized model setups (\Autoref{subseec:finalmodel}).

\subsection{Data augmentation}
\label{subsec:dataaug}
We implemented up-sampling augmentation to ensure balance exists at the class and subtype level in the training set (see \Autoref{apx:spectra-distribution-splits}). Also, a small number of augmented copies are allowed into the validation sets only for class 1, to improve subtype representation given the scarcity of unique objects in this class (\Autoref{tab:full-table-data-set}). \Autoref{fig:proportion_split} shows the subtype composition of the validation splits for each split and class. 

Two data augmentation techniques were applied to each upsampled copy in our upsampled data while avoiding the introduction of identical copies. Recalling that the dataset is comprised of spectra corrected to rest-frame redshift ($z$=0), in order to build robustness around wavelength calibration uncertainties, augmented spectra are randomly shifted by between 1 and 24 bins in wavelength space in either direction. Given the log-uniform sampling of 1,024 bins over [2,501.69 -- 9,993.24]~\AA~for all spectra, this corresponds to relative wavelength shifts of $\approx$0.14\%--3.3\% (or $\approx$ 406--9900~km\,s$^{-1}$), spanning the range from typical wavelength-calibration uncertainties to shifts comparable to small residual redshift errors. 
\begin{figure}
    \centering
    \includegraphics[width=\linewidth]{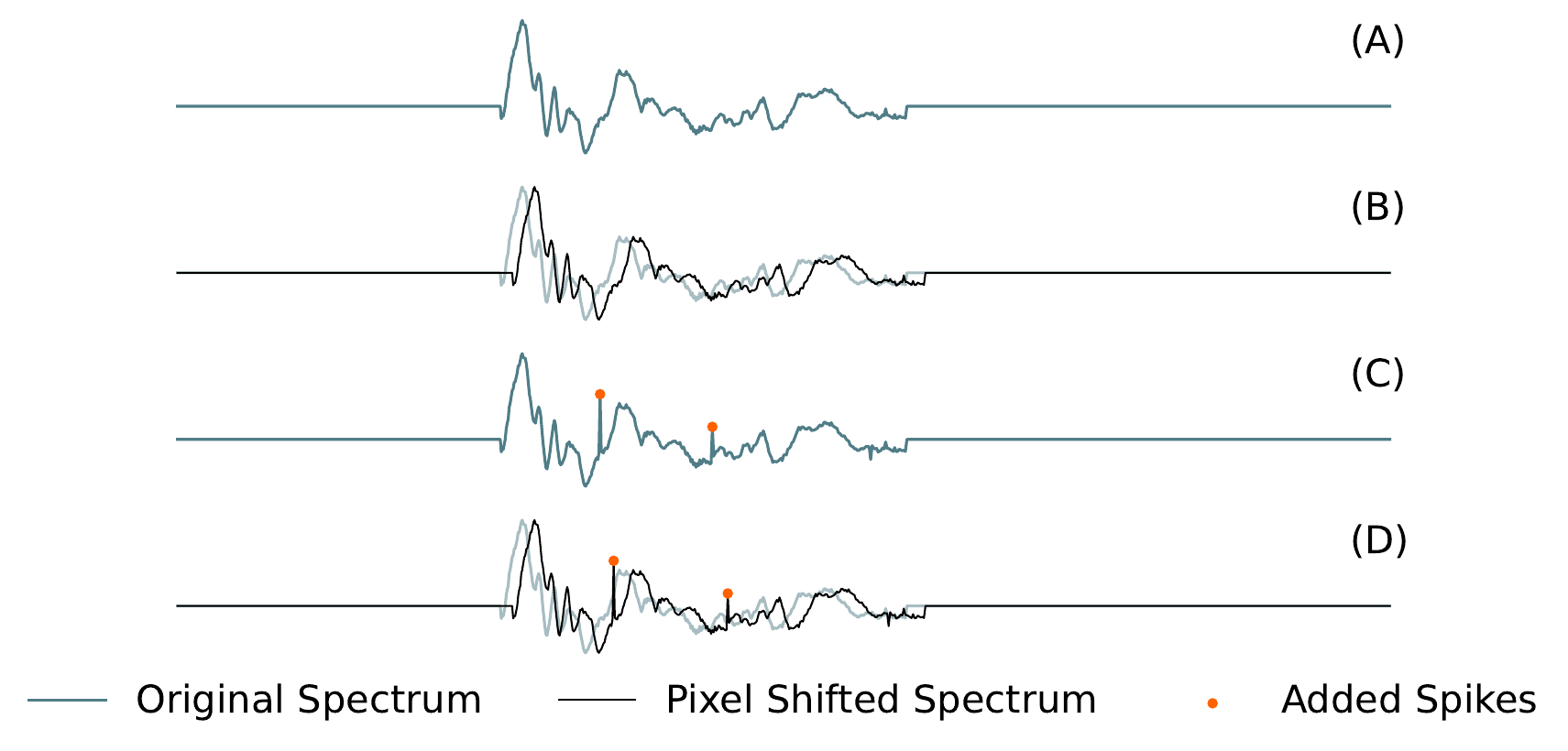}
    \caption{Spectrum of type Ia SN 2008A at a spectral phase of 0.791 days from peak brightness, from the validation dataset. Panel (A) shows the original, rest-frame spectrum. Panel (B) shows the pixel-shifted version of the spectrum with the original version in light turquoise for reference. Panel (C) shows a third possible augmentation of the spectrum, namely the addition of spikes. Panel (D) shows the last possible augmentation in which pixel shift and spike augmentations are both applied (original spectrum shown in light turquoise for reference).}
    \label{fig:treatment}
\end{figure}
A common artifact found in SN spectra is the presence of spikes, resembling those caused by telluric contamination, cosmic rays, or instrumental artifacts. Thus, we applied a second augmentation technique, adding artificial spikes that imitate the features already present in the dataset. Each spectrum was assigned a random number of spikes, chosen uniformly from the integers 0 to 5. Spike positions were determined as follows: the observed-frame wavelengths of three telluric features (H$_2$O at 590~nm, O$_2$ B-band at 690~nm, O$_2$ A-band at 760~nm) were mapped into the rest-frame spectrum using each SN's redshift, and a spike was placed at the nearest wavelength bin for each telluric feature that fell within the spectrum's valid (non-zero) wavelength range. Any remaining spikes were placed at randomly chosen wavelength bins. The amplitude of each spike was drawn from a half-normal distribution, and 80\% of the spikes were chosen to mimic an emission (positive) line (such as telluric lines or cosmic rays), and 20\% an absorption (negative) line, generally mimicking incorrect telluric line correction or other artifacts.

For each upsampled spectrum, the applied modification was sampled from a discrete uniform distribution among four options: no modification, pixel-shift only, spikes only, or spikes combined with pixel-shift. Examples of these possible modifications are illustrated in \Autoref{fig:treatment}. To prevent the network from associating spectral modifications with augmented copies specifically, the same treatments were randomly applied to a subset of the original Ia-norm spectra (the largest class, and hence not augmented). While a common augmentation technique involves increasing the noise in the input data, the noise characteristics of spectra are highly specific, depending on the intrinsic shot noise, as well as systematic instrumental response effects. Furthermore, astrophysical spectra are often reported in their already processed form without the associated noise arrays \citep[e.g.,][]{2016ApJ...827...90L}, thus we refrained from applying noise augmentations. However, we note that the underlying dataset already reflects the heterogeneity of real instruments, observing conditions, and reduction pipelines (\Autoref{subsec:balancing}), providing an implicit diversity of noise properties.  Nonetheless, this diversity does not substitute for targeted noise augmentation or establish robustness in lower-SNR regimes. Similarly, we did not include host-galaxy contamination in the augmentation scheme. Given the nature of the dataset (\Autoref{subsec:balancing}), it is not expected to contain enough examples with strong narrow host-galaxy emission lines to assess their effect on the flux scaling of the input images and on classification. Robustness to such contamination is therefore not established here, and could be addressed in future work by combining SN spectra with galaxy templates during training, as done in DASH \citep{2019DASH}.

\begin{figure}
    \centering
    \includegraphics[width=0.5\linewidth]{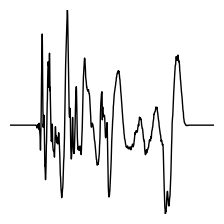}
    \caption{Sample spectral visualization used for training and testing, illustrating the spectrum of type Iax SN 2005kc at a spectral phase of 12.187 days. This figure is further discussed in \Autoref{sssec:spectra}.}
    \label{fig:sampleInput}
\end{figure}
 
\subsection{Data encoding}
\label{subsec:data_visualization}
As noted in \Autoref{sec:introduction}, one of the goals of this study is to assess how different visual representations of the same data influence the performance of ViTs–like models. For this purpose, two visualization approaches were considered: a direct y-vs-x (line-plot of flux vs. wavelength) visualization of the spectra and a \dfdw\ map representation. From each visualization, PNG images of size 224$\times$224 pixels were generated, corresponding to a figure size of 2.24$\times$2.24 inches at 100 dpi. This resolution matches the required input format for most of the pre-trained models used in this study. Both visualization methods were applied to the augmented dataset and are described in more detail below. 

\begin{figure*}
    \centering

    \begin{subfigure}{0.19\textwidth}
        \includegraphics[width=\linewidth]{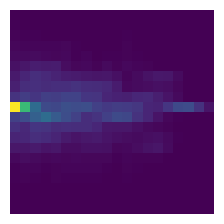}
        \caption*{20$\times$20 (natural scale)}
    \end{subfigure}
    \hfill
    \begin{subfigure}{0.19\textwidth}
        \includegraphics[width=\linewidth]{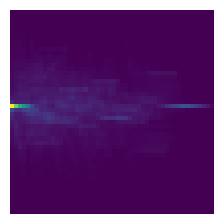}
        \caption*{50$\times$50 (natural scale)}
    \end{subfigure}
    \hfill
    \begin{subfigure}{0.19\textwidth}
        \includegraphics[width=\linewidth]{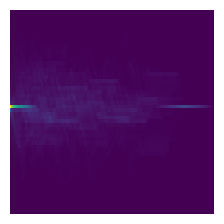}
        \caption*{224$\times$56 (natural scale)}
    \end{subfigure}
    \hfill
    \begin{subfigure}{0.19\textwidth}
        \includegraphics[width=\linewidth]{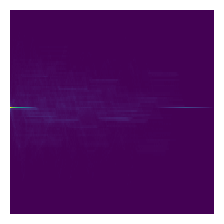}
        \caption*{224$\times$112 (natural scale)}
    \end{subfigure}
    \hfill
    \begin{subfigure}{0.19\textwidth}
        \includegraphics[width=\linewidth]{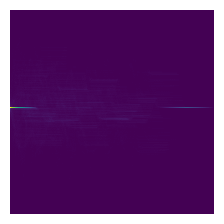}
        \caption*{224$\times$224 (natural scale)}
    \end{subfigure}

    \vspace{0.3cm}

    \begin{subfigure}{0.19\textwidth}
        \includegraphics[width=\linewidth]{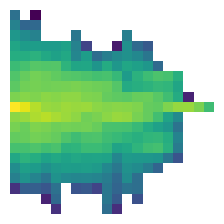}
        \caption*{20$\times$20 (log scale)}
    \end{subfigure}
    \hfill
    \begin{subfigure}{0.19\textwidth}
        \includegraphics[width=\linewidth]{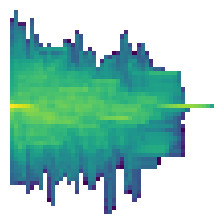}
        \caption*{50$\times$50 (log scale)}
    \end{subfigure}
    \hfill
    \begin{subfigure}{0.19\textwidth}
        \includegraphics[width=\linewidth]{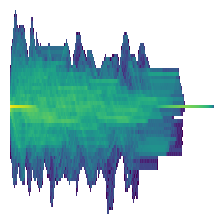}
        \caption*{224$\times$56 (log scale)}
    \end{subfigure}
    \hfill
    \begin{subfigure}{0.19\textwidth}
        \includegraphics[width=\linewidth]{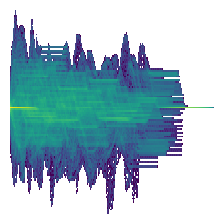}
        \caption*{224$\times$112 (log scale)}
    \end{subfigure}
    \hfill
    \begin{subfigure}{0.19\textwidth}
        \includegraphics[width=\linewidth]{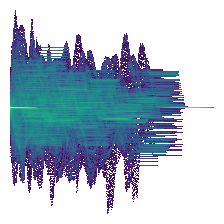}
        \caption*{224$\times$224 (log scale)}
    \end{subfigure}

    \caption{Examples of \dfdw\ map representations of the spectrum of type Iax SN 2005kc at a spectral phase of 12.187 days (also used in \Autoref{fig:sampleInput}), shown for different binning refinement levels and color scales. The top row displays maps in natural color scale, and the bottom row displays the same configurations in logarithmic color scale. From left to right, the binning resolutions are 20$\times$20, 50$\times$50, 224$\times$56, 224$\times$112, and 224$\times$224, illustrating the effect of binning refinement on the resulting image structure. This figure is further discussed in \Autoref{sssec:dfdw}.}
    \label{fig:maps}
\end{figure*}

\subsubsection{Flux-wavelength plot}
\label{sssec:spectra}
\Autoref{fig:sampleInput} illustrates the spectrum of SN 2005kc converted into a flux vs. wavelength line-plot. Hereafter, we refer to this as a \fwp. No axes, scales, or grid lines were included in the images to avoid introducing additional visual cues that could bias the network. This removes the information about overall energy emission in the SN, but in most cases the spectra are not available in physical units anyway.

By adopting this minimal representation, the model is encouraged to focus on recognizing spectral shapes that are characteristic of different SN types. This design choice also makes the input format independent of axis labels or other visual artifacts, and easily reproducible.

\subsubsection{\dfdw\ maps}
\label{sssec:dfdw}

We generate alternative representations of the data as  \dfdw\ maps, which encode differences in fluxes (\df) as a function of differences in wavelength (\dw). This approach is inspired by the \dm–\dt\ representation (where \dm\ denotes magnitude differences and \dt\  denotes time differences) used for light curves, as proposed by \citet{2017Mahabal} and later adopted by subsequent studies for training machine-learning models \citep{2021MG}. Consider that for each pair of points among the 1,024 wavelength-flux data points that comprise each spectrum, we determine the difference in \df\ and in \dw, giving $p=\binom{n}{2}=n(n-1)/2$ points for a spectrum of length $n$.  This representation is then a 2D histogram of the occurrence rate of $\Delta f$ and $\Delta w$ values, for a chosen binning size in both $f$ and $w$.
The advantage of the \dfdw\ representation is that it emphasizes spectral variability, potentially revealing features that may be less apparent in the direct spectral visualization. However, we note that this representation loses the information about a feature's wavelength location, which is the primary identifier for the chemical composition of SN ejecta.

Each spectrum consists of 1,024 wavelength bins, whereas the input images are fixed at 224$\times$224 pixels. As a result, rebinning is required, which also provides an opportunity to investigate the impact of binning refinement on classification performance. Five binning configurations were explored: 20$\times$20, 50$\times$50, 224$\times$56, 224$\times$112, and 224$\times$224. Because smaller magnitude differences are expected within a single source, the sensitivity of the models to the choice of color scaling was also investigated. Both natural and logarithmic color scales were tested. In total, ten distinct \dfdw\ datasets were generated, corresponding to the five binning schemes combined with the two color scales. These were used alongside the \fwp\ images described above.

\Autoref{fig:maps} presents examples of all generated \dfdw\ maps at the different binning levels and color scales, using the spectrum of SN 2005kc as a representative case.

\section{Experimental setup}
\label{sec:experimental_setup}
\subsection{Models} 
\label{subsec:models}
We adopted four Transformer-based vision backbones with different architectural inductive biases and pre-training strategies. We used the ViT architecture \citep{2020ViT}, which interprets an image as a sequence of fixed-size patches treated as tokens by a standard transformer encoder. Specifically, we employed the base-sized models pretrained on ImageNet-21k at 224$\times$224 resolution: \texttt{google/vit-base-patch32-224-in21k}\footnote{\url{https://huggingface.co/google/vit-base-patch32-224-in21k}} (\texttt{vitb-p32} for short) and \texttt{google/vit-base-patch16-224-in21k}\footnote{\url{https://huggingface.co/google/vit-base-patch16-224-in21k}} (\texttt{vitb-p16}). These two models share architecture but differ in the number of parameters ($\sim$88\,M vs. $\sim$86.4\,M) and patch size (32$\times$32 vs. 16$\times$16), with the smaller patch size providing a higher spatial token resolution and potentially finer sensitivity to localized spectral features, such as individual spectral lines in the flux–wavelength plots, or characteristic combinations of flux differences and wavelength separations in the \dfdw\ maps. Although pretrained on natural images, ViTs have demonstrated strong transfer performance due to their ability to model long-range dependencies through global self-attention, which is well suited to spectral classification where discriminative information is encoded in line positions, widths, and continuum shape \citep{2025FoxFortino}.

To assess the impact of alternative inductive biases and pre-training paradigms, we additionally considered two non-standard ViT variants. The \texttt{microsoft/swinv2-tiny-patch4-window16-256}\footnote{\url{https://huggingface.co/microsoft/swinv2-tiny-patch4-window16-256}} model (\texttt{swinv2t-p4}) is based on the Swin Transformer V2 architecture \citep{2022Liu}, which introduces a hierarchical structure and computes self-attention within local windows, enabling multi-scale feature representations and improved computational efficiency compared to global attention. This model was pretrained on ImageNet-1k at a resolution of $256\times256$ and contains approximately 27.6\,M parameters. Finally, we include the self-supervised \texttt{facebook/dinov3-vits16-pretrain-lvd1689m}\footnote{\url{https://huggingface.co/facebook/dinov3-vits16-pretrain-lvd1689m}} model (\texttt{dinov3-vits16}), a Vision Transformer with approximately 21.6\,M parameters, trained using the DINOv3 framework \citep{2025Simeoni} on a large unlabeled web-scale dataset (LVD-1689M). In DINOv3, smaller student models are trained via self-distillation from a substantially larger teacher network (ViT-7B). This distillation-based self-supervised training yields dense and transferable visual features, which may improve generalization in domains such as astronomical spectroscopy. DINOv3 is specifically optimized to function as a universal visual encoder without requiring task-specific fine-tuning. For each backbone, the corresponding HuggingFace image processor was used to resize, normalize, and format the input images according to the requirements of the pre-training procedure.

A common classification head was appended to the output of each backbone. This head consists of a three-layer multilayer perceptron (MLP): a fully connected layer projecting the backbone embeddings to 512 neurons, followed by a ReLU activation and dropout; a second fully connected layer with 256 neurons, also followed by ReLU and dropout; and a final linear layer mapping to the three target classes. The number of parameters of the classifier head varies with the size of the output feature vector from the backbones, which is 384 for \texttt{dinov3-vits16} ($\sim$0.33\,M parameters) and 768 for the rest ($\sim$0.53\,M parameters). A softmax activation was applied to the output layer to produce class probabilities. This unified classification head ensures that performance differences across experiments can be attributed to the backbone architectures and training strategies rather than to variations in the classifier design.

\subsection{Training and evaluation strategy} 
\label{subsec:training} 
The experimental strategy followed a staged evaluation procedure aimed at assessing the impact of (i) fine-tuning depth, (ii) visual representation of spectra, and (iii) model-specific hyperparameters, while accounting for data variability through multiple train/validation splits.

We used the AdamW \citep{2017Loshchilov} optimizer with weight decay for regularization and a linear warm-up of 4 epochs to stabilize optimization and avoid large gradient updates in the first epochs. The training objective was categorical cross-entropy loss, which is computed on the logits outputs. During training, we tracked the validation loss and macro-$F1$ score. Early stopping was triggered when validation loss did not improve for a certain number of epochs defined by a patience set for each experiment, and we restored weights to the epoch that achieved maximum macro-$F1$ score. The number of training epochs was set to 200; however, in all experiments, the early stopping criterion led to termination well before reaching this limit. To ensure reproducibility and to restrict variability to controlled sources, random seeds were fixed across all relevant stochastic components. The metric reported for each experiment is the mean and standard deviation validation macro-$F1$ score computed on three train/validation splits described in \Autoref{subsec:balancing}. All models were trained with an NVIDIA A100 GPU. 

\subsection{Effect of fine-tuning depth}
In the first stage, we investigated the impact of different fine-tuning strategies for the transformer backbone models. We considered three configurations: (i) using the pre-trained backbones as fixed feature extractors and training only the classifier head, (ii) partially fine-tuning the models by unfreezing the last four transformer layers, and (iii) fully fine-tuning all backbone layers. For this experiment, we fixed the spectral representation to the baseline \fwp\ (\Autoref{sssec:spectra}) and adopted a common set of well-behaved hyperparameters identified in preliminary experiments on this dataset. Specifically, the weight decay was set to \num{1e-4}, the batch size to 32, and the learning rate to \num{1e-3} for the frozen-backbone configuration, and to \num{1e-5} for both the partially and fully unfrozen configurations. The patience was set to 7 epochs (requiring any improvement, or \texttt{mindelta=0}). 

\begin{figure}
    \centering
    \includegraphics[width=1.0\linewidth]{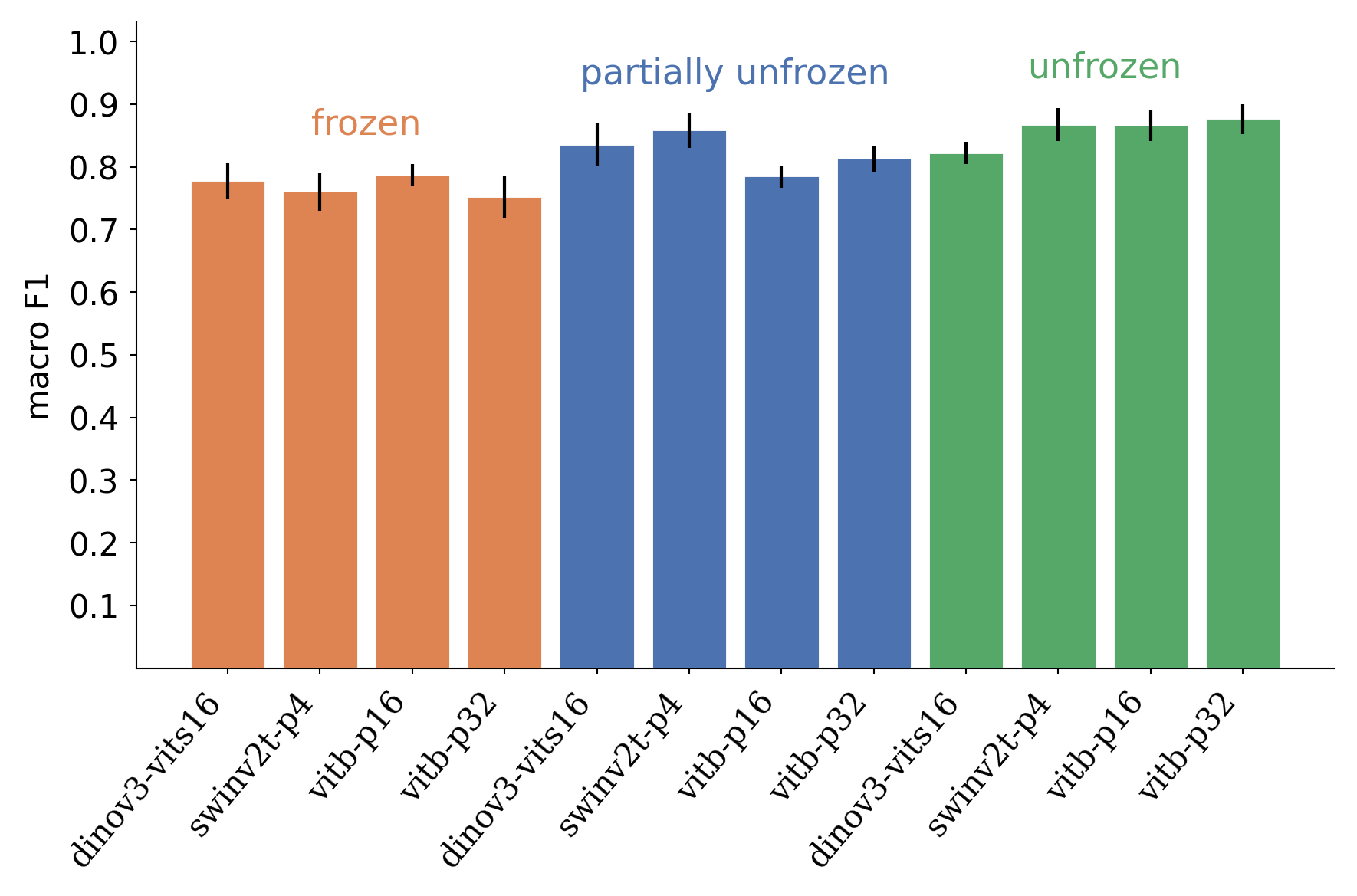}
    \caption{Mean macro-$F1$ scores for the four pre-trained models with different levels of fine-tuning for a fixed set of hyperparameters and visual representation of spectra. Error bars indicate the standard deviation across splits.}
    \label{fig:phase1}
\end{figure}

\Autoref{fig:phase1} summarizes the effect of different fine-tuning strategies across the four pre-trained backbones. Although fine-tuning is generally expected to improve performance, the magnitude of this effect is not a priori obvious in our setting, given the significant domain shift between natural images and spectral data. We observed that freezing the backbone consistently led to substantially lower predictive performance across all models. The gains from unfreezing, however, were not uniform and depended on both the fine-tuning depth and the backbone architecture. For \texttt{vitb-p16} and \texttt{vitb-p32}, partial fine-tuning yielded modest gains over the frozen regime, whereas full fine-tuning produced substantially larger improvements, reaching macro-$F1$ scores of $0.87\pm0.02$ and $0.88\pm0.02$. In contrast, \texttt{swinv2t-p4} benefits strongly from both levels of unfreezing, achieving $0.86\pm0.03$ with partial fine-tuning and $0.87\pm0.03$ with full fine-tuning. \texttt{dinov3-vits16} exhibits a distinct pattern: partial fine-tuning yielded a notable improvement over the frozen backbone ($0.84\pm0.03$ vs.\ $0.78\pm0.03$), but full fine-tuning did not further improve performance ($0.82\pm0.02$), suggesting that the deeper layers of this model may not benefit from additional adaptation in this domain. Notably, \texttt{dinov3-vits16} achieved the best performance among the frozen backbones, consistent with the findings of \citet{2025Simeoni}, who showed that the DINOv3 foundation model provides strong performance across a wide range of tasks even without fine-tuning. These performance differences come at a corresponding computational cost: freezing the backbone incurred the lowest per-epoch training time, while partially unfreezing the last four layers increased it by roughly 50--60\%, and fully unfreezing all layers raised it by a further factor of $\sim$2, resulting in per-epoch costs 2.5--2.8$\times$ higher than the frozen baseline. 

Based on these results, we adopted fully unfrozen backbones for all subsequent experiments, with the exception of \texttt{dinov3-vits16} and \texttt{swinv2t-p4}, for which we also retained the partially unfrozen configuration given its competitive performance at a lower computational cost.

\subsection{Impact of visual representations}
\label{sec:visual_representations}
In the second stage we aimed to explore the impact of using different visual encodings of the spectral data. We trained the six models using the same data and splits but with the 11 different visual representations: the base \fwp\ and the 10 variations of \dfdw\ maps (see \Autoref{subsec:data_visualization}). The hyperparameters and patience were the same used for the unfrozen and partially unfrozen models in the previous section. 

\begin{figure*}
    \centering
    \includegraphics[width=1.0\linewidth]{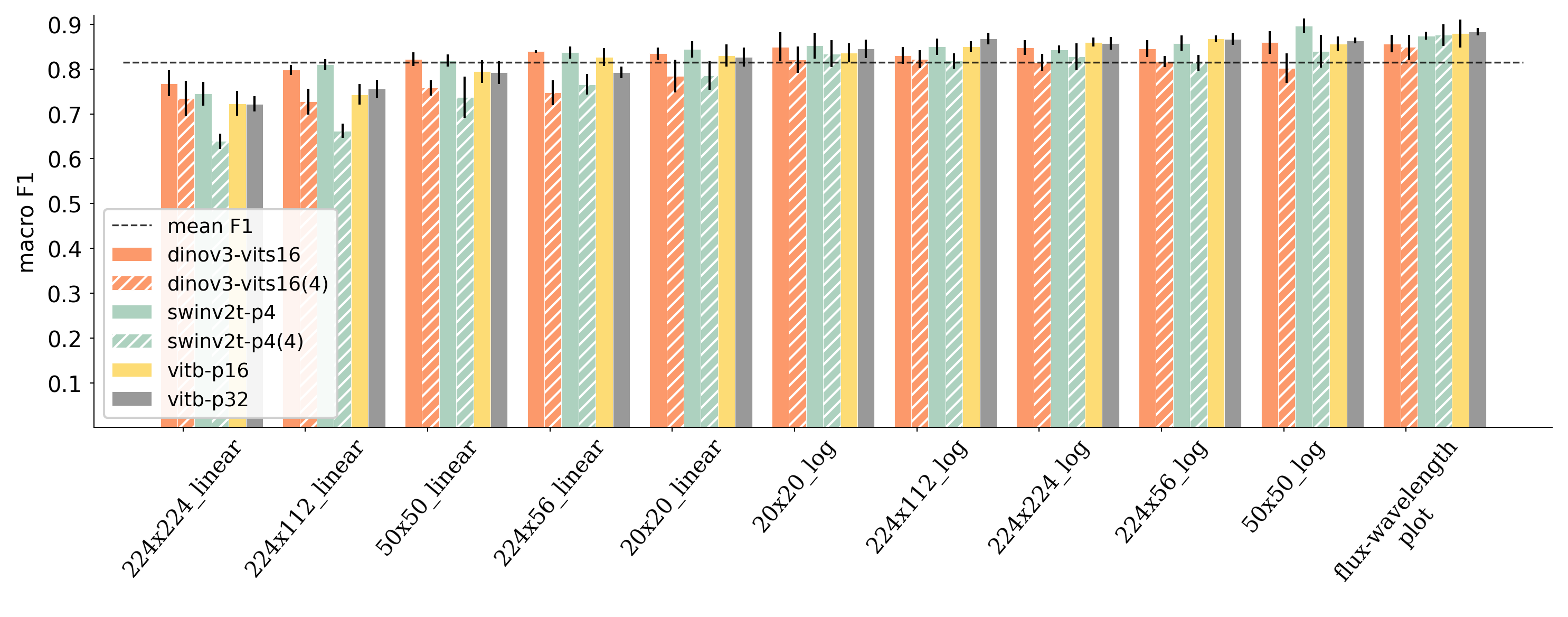}
    \caption{Mean macro-$F1$ scores for each model and visual representation. Error bars indicate the standard deviation across splits. Models labeled (4) correspond to the partially unfrozen fine-tuning regime.}
    \label{fig:phase2}
\end{figure*}

\Autoref{fig:phase2} summarizes the results. The \fwp\ ranks as the top-performing representation on average: across all six model configurations, its macro-$F1$ score lies at least $1\sigma$ above the mean computed over all models and representations. For \texttt{vitb-p16} and \texttt{vitb-p32}, the \fwp\ consistently achieves the highest macro-$F1$ scores ($0.88\pm0.03$ and $0.88\pm0.01$, respectively), outperforming all \dfdw\ map variants. For \texttt{swinv2t-p4} and \texttt{dinov3-vits16}, the outcome depends on the fine-tuning regime. For \texttt{dinov3-vits16} (the least performing backbone overall) the \fwp\ and \texttt{50x50\_log} map yielded comparable performance in both regimes, with the \fwp\ remaining competitive throughout. For \texttt{swinv2t-p4}, the \texttt{50x50\_log} map achieved the best macro-$F1$ across all models in the fully unfrozen regime ($0.90\pm0.02$ vs.\ $0.88\pm0.01$ for the \fwp), while in the partially unfrozen regime the \fwp\ clearly dominates ($0.88\pm0.02$ vs.\ $0.84\pm0.04$). Notably, representations in natural scale consistently underperformed their log-scale counterparts across all backbones and regimes, confirming the importance of logarithmic scaling for the \dfdw\ maps.

Based on these results, we proceeded as follows for the subsequent hyperparameter tuning stage: \texttt{vitb-p16} (fully unfrozen), \texttt{vitb-p32} (fully unfrozen), and \texttt{swinv2t-p4} (partially unfrozen) were carried forward using the \fwp\ representation only; \texttt{swinv2t-p4} (fully unfrozen) was carried forward with both the \fwp\ and the \texttt{50x50\_log} map, given the non-negligible performance advantage of the latter for this configuration.

\subsection{Hyperparameter optimization}
\label{subsec:hyperparam_optimization}
In the third stage, we optimized the hyperparameters for each of the selected models. We performed a grid search over a common set of hyperparameters spanning batch size, weight decay, and dropout rate. The learning rate was fixed to \num{1e-5} with a linear warm-up over four epochs, and early stopping was applied with a patience of 10 epochs.

\Autoref{table:fine_tuning} summarizes the explored search space and reports the best-performing hyperparameter configuration for each model.  Overall, all five configurations achieved comparable macro-$F1$ scores. With \texttt{swinv2t-p4} and the \texttt{50x50\_log} representation achieving the best overall performance ($0.900\pm0.008$), and also exhibiting the smallest cross-split variance among all configurations, followed closely by \texttt{swinv2t-p4} with the \fwp\ representation and \texttt{vitb-p32} ($0.892\pm0.018$ and $0.888\pm0.015$, respectively), while \texttt{vitb-p16} and \texttt{swinv2t-p4} in the partially unfrozen regime yielded somewhat lower macro-$F1$ scores ($0.879\pm0.012$ and $0.876\pm0.029$), with the latter also showing the largest cross-split variance.

\begin{table*}[h!]
\caption{Hyperparameter optimization results.}
\label{table:fine_tuning}
\centering
\begin{tabular}{l l c c c c}
\hline\hline
Model & Representation & Batch size & Weight decay & Dropout & macro-$F1$ \\
 & & \{32, 64\} & \{\num{1e-4}, \num{1e-5}\} & \{0.1, 0.3, 0.5\} & \\
\hline
\texttt{swinv2t-p4}  & 50x50\_log & 32 & \num{1e-5} & 0.5 & $0.900\pm0.008$ \\
\texttt{swinv2t-p4}  & flux-wavelength & 32 & \num{1e-5} & 0.1 & $0.892\pm0.018$ \\
\texttt{vitb-p32}    & flux-wavelength & 32 & \num{1e-5} & 0.3 & $0.888\pm0.015$ \\
\texttt{vitb-p16}   & flux-wavelength & 32 & \num{1e-5} & 0.1 & $0.879\pm0.012$ \\
\texttt{swinv2t-p4(4)} & flux-wavelength & 32 & \num{1e-4} & 0.3 & $0.876\pm0.029$ \\
\hline\hline
\end{tabular}
\tablefoot{For each model, the selected hyperparameters correspond to the best-performing configuration within the explored search space. Models labeled (4) correspond to the partially unfrozen fine-tuning regime.}
\end{table*}

\subsection{Test set evaluation}
\label{subseec:finalmodel}
Having selected the best hyperparameters for the five models, we performed the final training using the fourth train/validation split (\Autoref{tab:split4}), which was not used for the selection of fine-tuning depth, visual representation and hyperparameter search. We trained the models with early stopping at 10 epochs and kept the weights at the best epoch based on macro-$F1$.

We initially assessed the models' overall performance on the test set using class-averaged precision, recall, and macro-$F1$. \Autoref{table:test_results} summarizes the results. Among them, \texttt{vitb-p16} achieved the highest macro-$F1$ ($0.863$), outperforming all other configurations across all three metrics. The two \texttt{swinv2t-p4} fully unfrozen configurations follow closely and performed similarly to each other regardless of representation ($0.841$ with \texttt{50x50\_log} and $0.840$ with \fwp), while the partially unfrozen variant and \texttt{vitb-p32} yielded somewhat lower scores ($0.823$ and $0.814$, respectively). 

Comparing these results with the validation scores in \Autoref{table:fine_tuning} reveals a modest validation-to-test gap: all configurations score between 0.02 and 0.07 lower on the held-out test set than on validation, and the relative ranking of models is not preserved. In particular, \texttt{swinv2t-p4} with the \texttt{50x50\_log} representation, the top-ranked configuration on validation ($0.900\pm0.008$), drops to second place on the test set ($0.841$), whereas \texttt{vitb-p16} exhibits the smallest degradation ($0.879\pm0.012$ to $0.863$) and emerges as the best overall model. A gap of this magnitude is not unexpected: the validation splits were used at every model selection stage (fine-tuning depth, visual representation, and hyperparameter search), which can induce mild overfitting to the validation data through the selection process itself, and differences in the subtype and phase composition of the test set may contribute as well. 

To further characterize class-dependent behavior and error modes, we report per-class precision, recall, $F1$-score, and false positive rate (FPR) in \Autoref{table:per_class_metrics}, and show normalized confusion matrices and one-vs-rest precision--recall curves in \Autoref{fig:per-class-figs}. These results reveal consistent trends across architectures: with both \texttt{vitb-p16} and \texttt{swinv2t-p4} (50x50\_log) achieving the strongest and most balanced performances for the Ia-norm and Ia-other classes, while all models perform comparably on the CC class. Differences between models are most pronounced for Ia-other, where recall and false positive rates vary substantially across configurations, highlighting this as the most challenging class to identify reliably.

\begin{table}[h!]
\footnotesize
\caption{Overall test-set performance for all final models.}
\label{table:test_results}
\centering
\begin{tabular}{l c c c c}
\hline\hline
Model & Representation & Precision & Recall & macro-$F1$ \\
\hline
vitb-p16      & flux-wavelength & 0.864  &  0.871  &  0.863 \\
swinv2t-p4    & 50x50\_log      & 0.842  &  0.863  &  0.841 \\
swinv2t-p4    & flux-wavelength & 0.843  &  0.843  &  0.840 \\
swinv2t-p4(4) & flux-wavelength & 0.846  &  0.809  &  0.823 \\
vitb-p32      & flux-wavelength & 0.824  &  0.807  &  0.814 \\
\hline
\hline
\end{tabular}
\tablefoot{Metrics are averaged across classes. Models labeled (4) correspond to the partially unfrozen fine-tuning regime.}
\end{table}

\begin{table*}
\caption{Per-class test-set performance of the final models.}
\label{table:per_class_metrics}
\centering
\begin{tabular}{l c cccc cccc cccc}
\hline\hline
 & & \multicolumn{4}{c}{Ia-norm} & \multicolumn{4}{c}{Ia-other} & \multicolumn{4}{c}{CC} \\
Model & Representation & $P$ & $R$ & $F1$ & FPR & $P$ & $R$ & $F1$ & FPR & $P$ & $R$ & $F1$ & FPR \\
\hline
\texttt{vitb-p16} & flux-wavelength & \textbf{0.95} & 0.93 & \textbf{0.94} & 0.06 & 0.69 & 0.86 & \textbf{0.77} & 0.09 & 0.95 & 0.82 & 0.88 & 0.02 \\
\texttt{swinv2t-p4} & 50x50\_log & \textbf{0.95} & 0.83 & 0.89 & \textbf{0.05} & 0.59 & \textbf{0.88} & 0.71 & 0.15 & \textbf{0.98} & 0.88 & \textbf{0.93} & \textbf{0.01} \\
\texttt{swinv2t-p4} & flux-wavelength & 0.90 & 0.90 & 0.90 & 0.11 & 0.66 & 0.78 & 0.71 & 0.10 & 0.97 & 0.85 & 0.91 & \textbf{0.01} \\
\texttt{swinv2t-p4(4)} & flux-wavelength & 0.86 & \textbf{0.95} & 0.90 & 0.17 & \textbf{0.73} & 0.58 & 0.65 & \textbf{0.05} & 0.95 & \textbf{0.90} & 0.92 & 0.02 \\
\texttt{vitb-p32} & flux-wavelength & 0.88 & 0.93 & 0.90 & 0.14 & 0.64 & 0.67 & 0.65 & 0.09 & 0.96 & 0.83 & 0.89 & \textbf{0.01} \\
\hline\hline
\end{tabular}
\tablefoot{The metrics are
precision ($P$), recall ($R$), $F1$-score ($F1$), and false positive rate (FPR). The best score for each metric and class is highlighted.}
\end{table*}

\section{Discussion}
\label{sec:discussion}
To contextualize our work, we compare our metrics with published results for automated classifiers for SN spectra. 

Several machine–learning approaches for spectroscopic SN classification have been developed in the literature. Methods designed for binary discrimination between Type Ia and non–Ia SNe, such as SNIa-score \citep{2021Fremling}, emphasize purity for automatic Ia identification in live follow–up contexts. SNIa-score reports a false–positive rate below 0.6\% while correctly classifying 80–90\% of Ia spectra, but it is optimized on classifying spectra from one specific survey instrument. In contrast, we explored the capabilities of Vision Transformer models to classify spectra from a heterogeneous source dataset (albeit with high SNR), addressing a subtly more specific three-class classification problem. Our best-performing model achieves an overall precision of $P=86.4\,\%$, recall of $R=87.1\,\%$, and macro-$F1$ of $86.3\%$, rising to $F1=94\%$ for the Ia-norm class. These results are not directly comparable to those of binary classification approaches, but they provide context and suggest competitive performance.
A complementary approach was presented by \citet{Sharma2025}, who introduced CCSNscore, a hierarchical deep-learning framework for classifying core-collapse supernovae using ultra–low resolution spectra from the SED-machine (SEDM) spectrograph, optionally supplemented with photometry. CCSNscore achieves high accuracy for broad CCSN typing (Type II versus Type Ibc), but, as SNIa-score, it is tailored to a specific instrument and resolution regime. We did not split the CC subclasses so we cannot perform a comparison here.

Multi–class models include DASH \citep{2019DASH}, the more recent ABC-SN \citep{2025FoxFortino}, and SpectraNet \citep{xu2026applecider} frameworks. DASH employs convolutional neural networks to classify into 16 subtypes and reports agreement of $\sim$93\% with human classifications on OzDES (Australian Dark Energy Survey; \citealt{2017MNRAS.472..273C}), although on a test set heavily biased towards the Ia-norm type. ABC-SN extends this approach with attention mechanisms, achieving an overall accuracy of 85\%, recalls $\ge$73\% across 10 subtypes, and a precision of $\sim$95\% on Ia-norm spectra, outperforming an updated and retrained DASH baseline on almost all fronts. SpectraNet further builds on the application of convolutional neural networks to the SN spectra classification problem by employing a multiscale approach, and achieves performance comparable to ABC-SN. We focus our comparison on ABC-SN and DASH. We inherited our training data from ABC-SN, although we made some modifications in the augmentation strategy, and we note that \citealt{fortino2026low} also systematically retrained DASH to enable a direct comparison between the two models. Our performance metrics are broadly in line  with ABC-SN: though direct numerical comparisons are limited by differences in taxonomies, data treatments, and evaluation protocols, our own macro-$F1$ (86.3\%) is very close to the derived macro-$F1$ for ABC-SN (83.01\%), and correspondingly outperforms DASH.

We did not perform a matched comparison with template-matching tools such as SNID, which is used as ground truth in our work, as in \citep{fortino2026low} and \citep{2019Muthukrishna}. Thus, neither superiority nor inferiority is established: our goal is not to establish whether Transformer-based classification is competitive with operational workflows used by the community, but to compare candidate off-the-shelf vision-Transformer architectures and spectral encodings to explore pathways of research in that direction. A comparison with human-assisted workflows would require a defined dataset, taxonomy, metrics, and permitted human involvement. For evaluations of an automated classifier with template-matching tools, we refer the reader to \citet{Kim2024} and \citet{milligan2025testing}.\footnote {Although we note that the former applies DASH to low-resolution spectra without retraining, thus a domain shift should be expected to impact performance, and the latter is a study based on simulated data.}

\begin{figure*}[h!]
    \centering
    \includegraphics[width=0.9\linewidth]{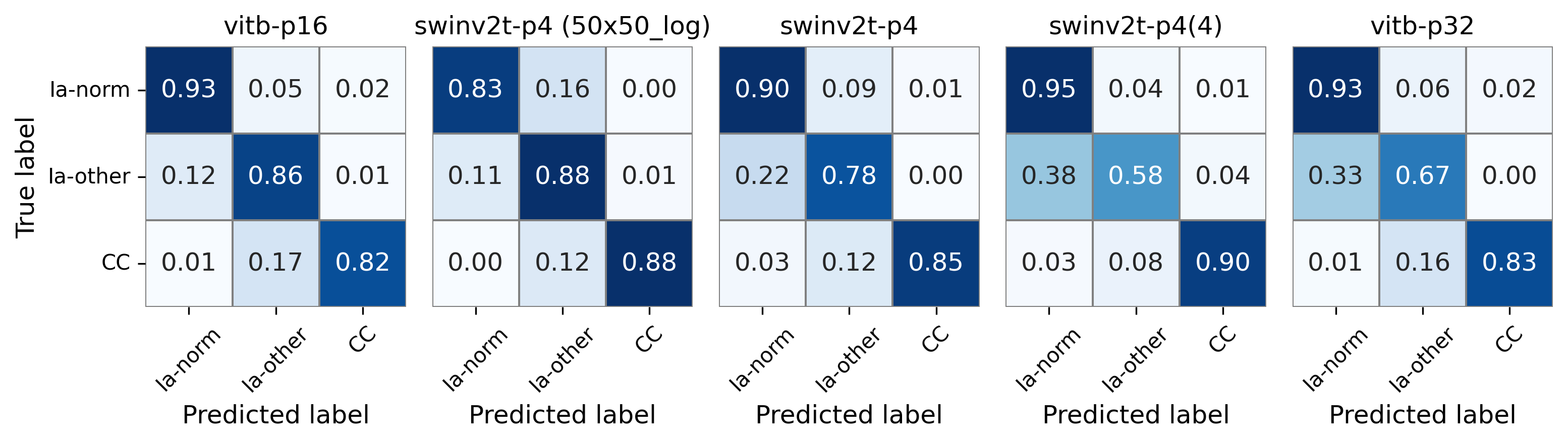}
    \includegraphics[width=0.9\linewidth]{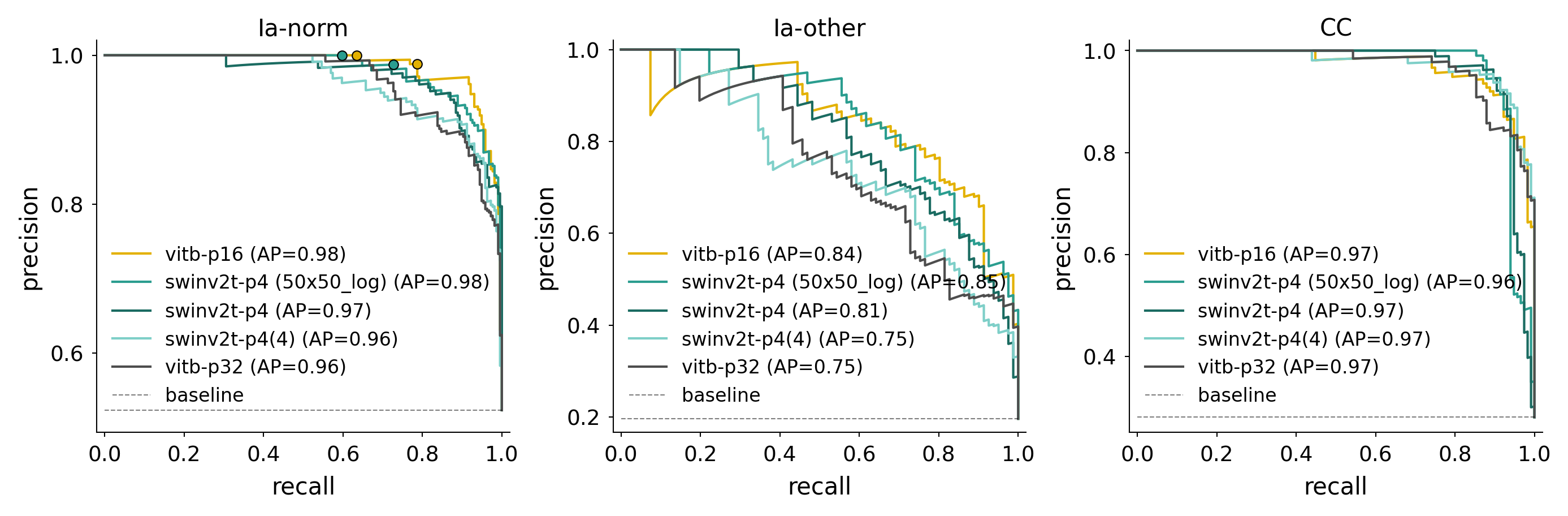}

    \caption{Held-out test-set confusion matrices and class-wise precision–recall performance for the five final models. Top panel: Confusion matrices normalized by true class counts. Bottom panel: One-vs-rest precision–recall curves for each class, with one panel per class and curves shown for all five models. The horizontal dashed line in each panel indicates the no-skill baseline, equal to the class prevalence in the test set, which differs across panels due to class imbalance. Average precision ($AP$), annotated in each panel, measures the area under the precision–recall curve and summarizes performance across all classification thresholds. In the Ia-norm panel, filled dots mark the operating points at 98.4\% and 100\% purity for \texttt{vitb-p16} ($78.7\%$ and $63.4\%$ recall, respectively) and \texttt{swinv2t-p4} with \texttt{50x50\_log} (72.7\% and 60.0\% recall, respectively).}
    \label{fig:per-class-figs}
\end{figure*}

In our work, a systematic misclassification pattern was observed throughout the experiments: Ia-other was consistently the hardest class to identify reliably, with errors occurring in both directions. The dominant confusion was the misclassification of Ia-other spectra as Ia-norm, with fractions ranging from 11\% in the best-performing models (\texttt{vitb-p16} and \texttt{swinv2t-p4} with \texttt{50x50\_log}) to 38\% in the partially unfrozen \texttt{swinv2t-p4(4)} (\Autoref{fig:per-class-figs}).  However, we remark that this is not a SN class per-se, but a collection of different classes with similar but distinct presentations (as is class 2), characterized by their similarity to SNe Ia. A secondary but consistent confusion was observed in the opposite direction, with 12--17\% of CC spectra misclassified as Ia-other across most models. Close inspection of the misclassified Ia-other samples reveals that the confusion is predominantly driven by the Ia-91T subtype. Type~Ia-91T events are known to be spectroscopically peculiar primarily at early phases, exhibiting weak or absent intermediate-mass element absorption features and prominent Fe~III lines, while at later epochs their spectra evolve to closely resemble those of normal SNe~Ia \citep{1992Phillips,2024Phillips}, making them among the most challenging to separate from the bulk Ia population. This phase dependence is clearly reflected in the pooled misclassification rates across all models: only 23.3\% of pre-maximum Ia-91T spectra (phase~$<0$) are misclassified, compared to 62.5\% at post-maximum phases (phase~$\geq 0$), consistent with the expectation that later-phase Ia-91T spectra become increasingly indistinguishable from normal Ia. The same trend is reported by other computer-vision classifiers: both ABC-SN and DASH identify Ia-91T events as the most frequently confused with normal Ia. The persistence of this confusion across independent models and architectures suggests that improved subtype discrimination for Ia-91T remains an open challenge for spectroscopic classifiers.

As for all predictive models, our absolute performance and relative rankings apply to the evaluation performed here and may change under distribution shift. Independent observational validation would help assess out-of-distribution generalization. Operational development would additionally require representative training coverage, appropriate augmentation, validation across observing conditions, and monitoring of distribution shifts and performance.

While our primary focus is classification performance, computational complexity is also an important consideration when comparing large pre-trained ViT architectures. The evaluated models span a wide range of parameter counts, from 27.6M parameters for \texttt{swinv2t-p4} to approximately 88M for the \texttt{vitb} variants. Despite having roughly three times fewer parameters, \texttt{swinv2t-p4} incurred a per-epoch training cost comparable to \texttt{vitb-p16} ($\sim$42--43\,s per epoch for both), owing to its shifted-window attention mechanism which, while linear in sequence length, operates over a larger number of hierarchical stages. By contrast, \texttt{vitb-p32} is the least expensive model at $\sim$19\,s per epoch, a direct consequence of its larger patch size reducing the token sequence length. The partially unfrozen \texttt{swinv2t-p4} configuration offers a middle ground at $\sim$27\,s per epoch, achieving competitive performance ($F1=0.823$) at roughly 60\% of the cost of its fully unfrozen counterpart. Notably, \texttt{vitb-p16} achieved the best overall test-set performance ($F1=0.863$) at a per-epoch cost comparable to \texttt{swinv2t-p4}, suggesting that for this task the inductive biases of the hierarchical Swin architecture did not translate into a decisive advantage over the isotropic ViT architecture.

\section{Conclusions}
\label{sec:conclusion}
In this work, we explored the potential of pre-trained transformer-based vision models for the spectral classification of SNe into three classes: Ia normal, other Ia subtypes and core-collapse supernovae. We conducted a staged evaluation of different fine-tuning strategies, architectures and 2D representations of 1D spectra, carrying forward five final model configurations to test-set evaluation.

We found that model performance improved significantly when the backbone was fully fine-tuned compared to classification based on frozen representations, while partial fine-tuning can also yield substantial gains at a lower computational cost. This indicates that although pre-training on natural images provides a strong initialization for feature extraction, deeper adaptation of the learned representations is required to bridge the domain shift to astronomical spectroscopy.

The \fwp\ representation consistently outperformed the \dfdw\ maps on average across all six model configurations. We ascribe this to the \dfdw\ encoding losing the information about the wavelength location of spectral features, which is the primary identifier for the chemical composition of SN ejecta (\Autoref{sssec:dfdw}). Nevertheless, \dfdw\ maps remained competitive for specific architecture and fine-tuning combinations: the \texttt{50x50\_log} map paired with the fully unfrozen \texttt{swinv2t-p4} achieved the best single-model validation result overall and the second best result in held-out data, suggesting that alternative encodings that emphasize spectral variability can be advantageous. Among \dfdw\ variants, log-scale maps consistently outperformed their natural-scale counterparts across all models and binning configurations, consistent with the expectation that logarithmic scaling better captures the dynamic range of small flux differences within a single spectrum (see \Autoref{sec:visual_representations}).

Among the evaluated configurations, \texttt{vitb-p16} achieved the best overall test-set performance with a macro-$F1$ of 86.3\% and per-class $F1$ scores of 94\% for Ia-norm, 77\% for Ia-other, and 88\% for CC. 
The purity of the SN Ia sample can be further enhanced by choosing a class probability threshold of $p>0.986$ ($p>0.994$) leading to 98.4\% (100\%) purity and 78.7\% (63.4\%) recall (locations marked by filled dots in \Autoref{fig:per-class-figs}). 

The dominant and most persistent source of classification error was the confusion between Ia-91T and Ia-norm spectra, which occurred across all models and architectures. This confusion is strongly phase-dependent, consistent with the known spectral evolution of this subtype towards normal Ia at later epochs. A secondary confusion between CC and Ia-other was also consistently observed across models. Improved subtype discrimination for Ia-91T remains an open challenge for spectroscopic classifiers.

While our tests and benchmarks are limited to a simplified classification schema and to a specific dataset, the performance metrics are  broadly in line with published results for leading automated classifiers, especially ABC-SN, and are an improvement over DASH. A direct comparison with these two models is enabled by the fact that these are trained on the same root dataset, but differences in dataset taxonomies and augmentation limit the comparison. In this contextual sense, this study demonstrates that Vision Transformer architectures can achieve competitive performance for automated SN spectral classification from a single spectrum. While deploying a ViT classifier that can replace existing tools is outside of the scope of this work and would require further refinements and testing on unseen data outside of the SNID templates dataset, 
our work provides methodological guidance for research exploring vision-Transformer-based approaches to SN spectral classification, identifying directions for future refinements of ViT training strategies and data encodings to maximize the performance of ViT for SN classification.

\begin{acknowledgements}
We thank MJA for facilitating access to the resources needed to generate the images. This support was provided through the ANID FONDECYT Iniciación 11251912.

J.S.B., P.G.M., V.C.R., and M.R. would like to warmly thank the local organizing committee, the scientific committee, and all speakers and participants of the 2024 La Serena School of Data Science, held at the AURA Campus in La Serena, Chile. This work originated as a group project developed during the school. Thanks to funding from NSF (Award Number (FAIN): 1637359) applications from students in US institutions are eligible for full scholarships that cover all school expenses.

\end{acknowledgements}

\bibliographystyle{aa} 
\bibliography{aa} 

\onecolumn
\begin{appendix}
\nolinenumbers
\section{Spectra distribution in train, validation, and test sets per split}
\label{apx:spectra-distribution-splits}
\begin{figure}
    \centering
    \includegraphics[width=0.9\linewidth]{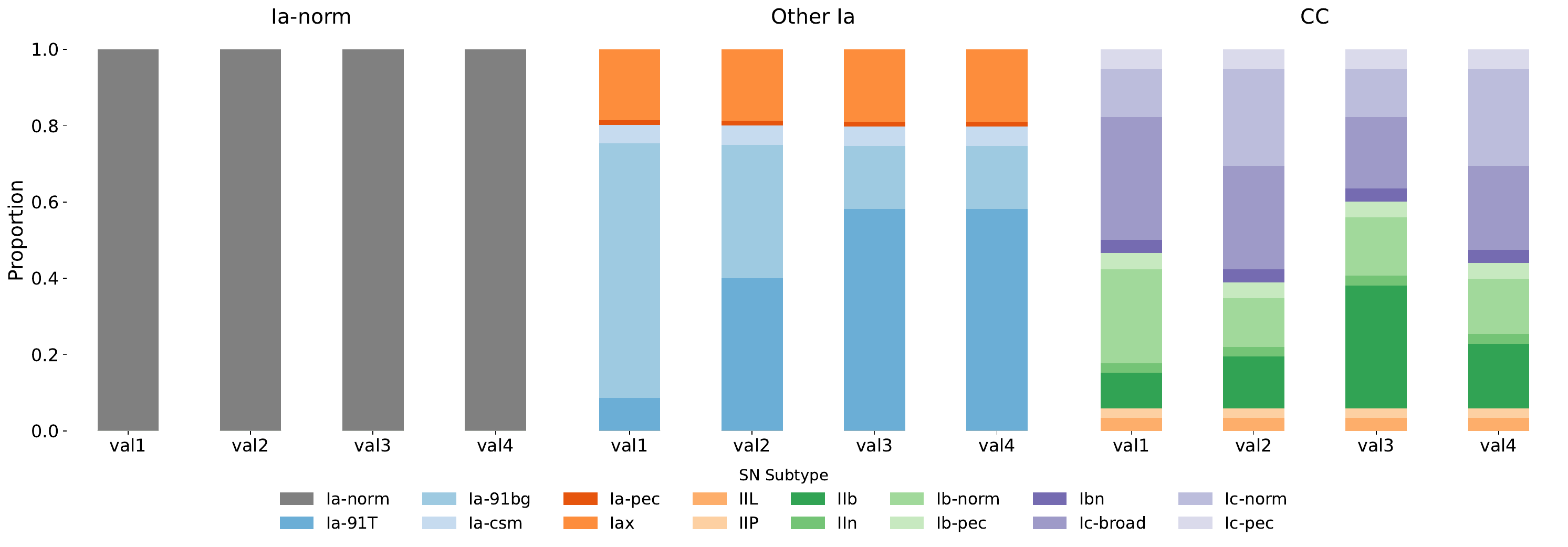}
    \caption{For each class and validation set (1, 2, 3, and 4), the proportion of each set that corresponds to each of the subtypes for that class.}
    \label{fig:proportion_split}
\end{figure}

This appendix presents the subtype-level distribution of spectra across the training, validation, and test sets for the four splits, as discussed in \Autoref{sec:data_setgen}. For each split, we report the total number of spectra, augmented copies, and unique SNe per class and subtype, broken down by set. \Autoref{fig:proportion_split} summarizes this information visually, showing the proportion of each validation set corresponding to each subtype.

\begin{table}[!htbp]
\footnotesize
\centering
\caption{}
\caption{Subtype-level statistics for Split~1.}
\label{tab:split1}
\begin{tabular}{c l | c c c | c c c | c c c}
\toprule\toprule
Class & Subtype &
\multicolumn{3}{c|}{Training} &
\multicolumn{3}{c|}{Validation} &
\multicolumn{3}{c}{Test} \\
\cmidrule(lr){3-5} \cmidrule(lr){6-8} \cmidrule(lr){9-11}
& &
\shortstack{Total\\Spectra} & \shortstack{Augmented\\Copies} & \shortstack{Unique\\SNe} &
\shortstack{Total\\Spectra} & \shortstack{Augmented\\Copies} & \shortstack{Unique\\SNe} &
\shortstack{Total\\Spectra} & \shortstack{Augmented\\Copies} & \shortstack{Unique\\SNe} \\
\midrule
\multirow{1}{*}{0}
& Ia-norm  & 1682 & 0   & 190 & 216 & 0  & 27 & 216 & 0 & 28 \\
\midrule
\multirow{5}{*}{1}
& Ia-91T   & 345  & 35  & 23  & 7   & 3  & 1  & 34  & 0 & 3  \\
& Ia-91bg  & 298  & 125 & 19  & 54  & 28 & 4  & 33  & 0 & 5  \\
& Ia-csm   & 352  & 340 & 1   & 4   & 0  & 1  & 0   & 0 & 0  \\
& Ia-pec   & 352  & 242 & 5   & 1   & 0  & 1  & 0   & 0 & 0  \\
& Iax      & 352  & 319 & 1   & 15  & 0  & 1  & 14  & 0 & 1  \\
\midrule
\multirow{11}{*}{2}
& II-pec   & 144  & 97  & 1   & 0   & 0  & 0  & 0   & 0 & 0  \\
& IIL      & 144  & 138 & 1   & 4   & 0  & 1  & 0   & 0 & 0  \\
& IIP      & 144  & 62  & 4   & 3   & 0  & 1  & 19  & 0 & 1  \\
& IIb      & 213  & 0   & 16  & 11  & 0  & 1  & 8   & 0 & 2  \\
& IIn      & 144  & 125 & 1   & 3   & 0  & 1  & 0   & 0 & 0  \\
& Ib-norm  & 159  & 0   & 16  & 29  & 0  & 2  & 23  & 0 & 4  \\
& Ib-pec   & 144  & 137 & 1   & 5   & 0  & 1  & 0   & 0 & 0  \\
& Ibn      & 144  & 139 & 1   & 4   & 0  & 1  & 18  & 0 & 1  \\
& Ic-broad & 188  & 0   & 18  & 38  & 0  & 3  & 2   & 0 & 2  \\
& Ic-norm  & 143  & 0   & 16  & 15  & 0  & 1  & 46  & 0 & 3  \\
& Ic-pec   & 144  & 119 & 1   & 6   & 0  & 1  & 0   & 0 & 0  \\
\bottomrule
\end{tabular}
\tablefoot{For each class and subtype, the total number of spectra, augmented copies, and unique SNe are reported separately for the training, validation, and test sets.}
\end{table}

\begin{table}[!htbp]
\footnotesize
\centering
\caption{Subtype-level statistics for Split~2.}
\label{tab:split2}
\begin{tabular}{c l | c c c | c c c | c c c}
\toprule\toprule
Class & Subtype &
\multicolumn{3}{c|}{Training} &
\multicolumn{3}{c|}{Validation} &
\multicolumn{3}{c}{Test} \\
\cmidrule(lr){3-5} \cmidrule(lr){6-8} \cmidrule(lr){9-11}
& &
\shortstack{Total\\Spectra} & \shortstack{Augmented\\Copies} & \shortstack{Unique\\SNe} &
\shortstack{Total\\Spectra} & \shortstack{Augmented\\Copies} & \shortstack{Unique\\SNe} &
\shortstack{Total\\Spectra} & \shortstack{Augmented\\Copies} & \shortstack{Unique\\SNe} \\
\midrule
\multirow{1}{*}{0}
& Ia-norm  & 1682 & 0   & 191 & 216 & 0  & 26 & 216 & 0 & 28 \\
\midrule
\multirow{5}{*}{1}
& Ia-91T   & 306  & 25  & 19  & 46  & 13 & 5  & 34  & 0 & 3  \\
& Ia-91bg  & 339  & 141 & 22  & 13  & 12 & 1  & 33  & 0 & 5  \\
& Ia-csm   & 352  & 340 & 1   & 4   & 0  & 1  & 0   & 0 & 0  \\
& Ia-pec   & 352  & 242 & 5   & 1   & 0  & 1  & 0   & 0 & 0  \\
& Iax      & 352  & 319 & 1   & 15  & 0  & 1  & 14  & 0 & 1  \\
\midrule
\multirow{11}{*}{2}
& II-pec   & 144  & 97  & 1   & 0   & 0  & 0  & 0   & 0 & 0  \\
& IIL      & 144  & 138 & 1   & 4   & 0  & 1  & 0   & 0 & 0  \\
& IIP      & 144  & 62  & 4   & 3   & 0  & 1  & 19  & 0 & 1  \\
& IIb      & 204  & 0   & 15  & 20  & 0  & 2  & 8   & 0 & 2  \\
& IIn      & 144  & 125 & 1   & 3   & 0  & 1  & 0   & 0 & 0  \\
& Ib-norm  & 171  & 0   & 17  & 17  & 0  & 1  & 23  & 0 & 4  \\
& Ib-pec   & 144  & 137 & 1   & 5   & 0  & 1  & 0   & 0 & 0  \\
& Ibn      & 144  & 139 & 1   & 4   & 0  & 1  & 18  & 0 & 1  \\
& Ic-broad & 200  & 0   & 19  & 26  & 0  & 2  & 2   & 0 & 2  \\
& Ic-norm  & 128  & 0   & 15  & 30  & 0  & 2  & 46  & 0 & 3  \\
& Ic-pec   & 144  & 119 & 1   & 6   & 0  & 1  & 0   & 0 & 0  \\
\bottomrule
\end{tabular}
\tablefoot{For each class and subtype, the total number of spectra, augmented copies, and unique SNe are reported separately for the training, validation, and test sets.}
\end{table}

\begin{table}[!htbp]
\footnotesize
\centering
\caption{Subtype-level statistics for Split~3.}
\label{tab:split3}
\begin{tabular}{c l | c c c | c c c | c c c}
\toprule\toprule
Class & Subtype &
\multicolumn{3}{c|}{Training} &
\multicolumn{3}{c|}{Validation} &
\multicolumn{3}{c}{Test} \\
\cmidrule(lr){3-5} \cmidrule(lr){6-8} \cmidrule(lr){9-11}
& &
\shortstack{Total\\Spectra} & \shortstack{Augmented\\Copies} & \shortstack{Unique\\SNe} &
\shortstack{Total\\Spectra} & \shortstack{Augmented\\Copies} & \shortstack{Unique\\SNe} &
\shortstack{Total\\Spectra} & \shortstack{Augmented\\Copies} & \shortstack{Unique\\SNe} \\
\midrule
\multirow{1}{*}{0}
& Ia-norm  & 1682 & 0   & 189 & 216 & 0  & 28 & 216 & 0 & 28 \\
\midrule
\multirow{5}{*}{1}
& Ia-91T   & 306  & 30  & 20  & 46  & 8  & 4  & 34  & 0 & 3  \\
& Ia-91bg  & 339  & 146 & 22  & 13  & 7  & 1  & 33  & 0 & 5  \\
& Ia-csm   & 352  & 340 & 1   & 4   & 0  & 1  & 0   & 0 & 0  \\
& Ia-pec   & 352  & 242 & 5   & 1   & 0  & 1  & 0   & 0 & 0  \\
& Iax      & 352  & 319 & 1   & 15  & 0  & 1  & 14  & 0 & 1  \\
\midrule
\multirow{11}{*}{2}
& II-pec   & 144  & 97  & 1   & 0   & 0  & 0  & 0   & 0 & 0  \\
& IIL      & 144  & 138 & 1   & 4   & 0  & 1  & 0   & 0 & 0  \\
& IIP      & 144  & 62  & 4   & 3   & 0  & 1  & 19  & 0 & 1  \\
& IIb      & 186  & 0   & 14  & 38  & 0  & 3  & 8   & 0 & 2  \\
& IIn      & 144  & 125 & 1   & 3   & 0  & 1  & 0   & 0 & 0  \\
& Ib-norm  & 170  & 0   & 16  & 18  & 0  & 2  & 23  & 0 & 4  \\
& Ib-pec   & 144  & 137 & 1   & 5   & 0  & 1  & 0   & 0 & 0  \\
& Ibn      & 144  & 139 & 1   & 4   & 0  & 1  & 18  & 0 & 1  \\
& Ic-broad & 204  & 0   & 20  & 22  & 0  & 1  & 2   & 0 & 2  \\
& Ic-norm  & 143  & 0   & 16  & 15  & 0  & 1  & 46  & 0 & 3  \\
& Ic-pec   & 144  & 119 & 1   & 6   & 0  & 1  & 0   & 0 & 0  \\
\bottomrule
\end{tabular}
\tablefoot{For each class and subtype, the total number of spectra, augmented copies, and unique SNe are reported separately for the training, validation, and test sets.}
\end{table}

\begin{table}[!htbp]
\footnotesize
\centering
\caption{Subtype-level statistics for Split~4.}
\label{tab:split4}
\begin{tabular}{c l | c c c | c c c | c c c}
\toprule\toprule
Class & Subtype &
\multicolumn{3}{c|}{Training} &
\multicolumn{3}{c|}{Validation} &
\multicolumn{3}{c}{Test} \\
\cmidrule(lr){3-5} \cmidrule(lr){6-8} \cmidrule(lr){9-11}
& &
\shortstack{Total\\Spectra} & \shortstack{Augmented\\Copies} & \shortstack{Unique\\SNe} &
\shortstack{Total\\Spectra} & \shortstack{Augmented\\Copies} & \shortstack{Unique\\SNe} &
\shortstack{Total\\Spectra} & \shortstack{Augmented\\Copies} & \shortstack{Unique\\SNe} \\
\midrule
\multirow{1}{*}{0}
& Ia-norm  & 1682 & 0   & 189 & 216 & 0  & 28 & 216 & 0 & 28 \\
\midrule
\multirow{5}{*}{1}
& Ia-91T   & 320  & 27  & 20  & 32  & 11 & 4  & 34  & 0 & 3  \\
& Ia-91bg  & 324  & 133 & 21  & 28  & 20 & 2  & 33  & 0 & 5  \\
& Ia-csm   & 352  & 340 & 1   & 4   & 0  & 1  & 0   & 0 & 0  \\
& Ia-pec   & 352  & 242 & 5   & 1   & 0  & 1  & 0   & 0 & 0  \\
& Iax      & 352  & 319 & 1   & 15  & 0  & 1  & 14  & 0 & 1  \\
\midrule
\multirow{11}{*}{2}
& II-pec   & 144  & 97  & 1   & 0   & 0  & 0  & 0   & 0 & 0  \\
& IIL      & 144  & 138 & 1   & 4   & 0  & 1  & 0   & 0 & 0  \\
& IIP      & 144  & 62  & 4   & 3   & 0  & 1  & 19  & 0 & 1  \\
& IIb      & 208  & 0   & 16  & 16  & 0  & 1  & 8   & 0 & 2  \\
& IIn      & 144  & 125 & 1   & 3   & 0  & 1  & 0   & 0 & 0  \\
& Ib-norm  & 173  & 0   & 17  & 15  & 0  & 1  & 23  & 0 & 4  \\
& Ib-pec   & 144  & 137 & 1   & 5   & 0  & 1  & 0   & 0 & 0  \\
& Ibn      & 144  & 139 & 1   & 4   & 0  & 1  & 18  & 0 & 1  \\
& Ic-broad & 194  & 0   & 19  & 32  & 0  & 2  & 2   & 0 & 2  \\
& Ic-norm  & 128  & 0   & 15  & 30  & 0  & 2  & 46  & 0 & 3  \\
& Ic-pec   & 144  & 119 & 1   & 6   & 0  & 1  & 0   & 0 & 0  \\
\bottomrule
\end{tabular}
\tablefoot{For each class and subtype, the total number of spectra, augmented copies, and unique SNe are reported separately for the training, validation, and test sets.}
\end{table}

\end{appendix}
\end{document}